\documentclass[prl,twocolumn,showpacs,preprintnumbers,amsmath,amssymb]{revtex4}

\usepackage{mathrsfs}
\usepackage{graphicx}
\usepackage{dcolumn}
\usepackage{bm}

\begin{document}

\title{Stochastic thermodynamic interpretation of information geometry}

\author{Sosuke Ito}
\affiliation{RIES, Hokkaido University, N20 W10, Kita-ku, Sapporo, Hokkaido 001-0020, Japan}
\date{\today}

\begin{abstract}
In recent years, the unified theory of information and thermodynamics has been intensively discussed in the context of stochastic thermodynamics. The unified theory reveals that information theory would be useful to understand non-stationary dynamics of systems far from equilibrium. In this letter, we have found a new link between stochastic thermodynamics and information theory well known as information geometry. By applying this link, an information geometric inequality can be interpreted as a thermodynamic uncertainty relationship between speed and thermodynamic cost. We have numerically applied an information geometric inequality to a thermodynamic model of biochemical enzyme reaction.
\end{abstract}


\maketitle
The crucial relationship between thermodynamics and information theory has been well studied in
last decades~\cite{parrondo2015thermodynamics}. Historically, thermodynamic-informational links had been discussed in the context of the second law of thermodynamics and the paradox of Maxwell's demon~\cite{leff2014maxwell}. Recently, several studies have newly revealed thermodynamic interpretations of informational quantities such as the Kullback-Leibler divergence~\cite{kawai2007dissipation}, mutual information~\cite{sagawa2010generalized, sagawa2012fluctuation, still2012thermodynamics}, the transfer entropy and information flow~\cite{ito2013information, hartich2014stochastic, hartich2016sensory, spinney2016transfer, ito2016backward, crooks2016marginal, allahverdyan2009thermodynamic, horowitz2014thermodynamics, horowitz2014second, shiraishi2015fluctuation, shiraishi2015role, yamamoto2016linear, goldt2017stochastic}. The above interpretations of informational quantities are based on the theory of stochastic thermodynamics~\cite{sekimoto2010stochastic, seifert2012stochastic}, which mainly focus on the entropy production in stochastic dynamics of small systems far from equilibrium. 

Information thermodynamic relationship has been attracted not only in terms of Maxwell's demon, but also in terms of geometry~\cite{crooks2007measuring,  feng2008length, polettini2013nonconvexity, shimazaki2015neurons, nicholson2015investigation, lahiri2016universal, tajima2017finite}. Indeed, differential geometric interpretations of thermodynamics have been discussed especially in a near-equilibrium system~\cite{weinhold1975metric, ruppeiner1979thermodynamics, salamon1983thermodynamic, crooks2007measuring, sivak2012thermodynamic, machta2015dissipation, rotskoff2017geometric}. Moreover, the technique of differential geometry in information theory, well known as information geometry~\cite{amari2007methods}, has received remarkable attention in the field of neuroscience, signal processing, quantum mechanics, and machine learning~\cite{amari2016information, pires2016generalized, oizumi2016unified}. In spite of the deep link between information and thermodynamics, the direct connection between thermodynamics and information geometry has been elusive especially for non-stationary and non-equilibrium dynamics. For example, G. E. Crooks discovered a link between thermodynamics and information geometry~\cite{crooks2007measuring, sivak2012thermodynamic} based on the Gibbs ensemble, and then his discussion is only valid for a near-equilibrium system.

In this letter, we discover a fundamental link between information geometry and thermodynamics based on stochastic thermodynamics for the master equation. We mainly report two inequalities derived thanks to information geometry, and interpret them within the theory of stochastic thermodynamics. The first inequality connects the environmental entropy change rate to the mean change of the local thermodynamic force rate.  The second inequality can be interpreted as a kind of thermodynamic uncertainty relationships or thermodynamic trade-off relationships~\cite{uffink1999thermodynamic, lan2012energy, govern2014optimal, ito2015maxwell, barato2015thermodynamic, gingrich2016dissipation, shiraishi2016universal, pietzonka2016universal, barato2016cost, horowitz2017proof, proesmans2017discrete, maes2017frenetic, dechant2017current} between speed of a transition from one state to another and thermodynamic cost related to the entropy change of thermal baths in a near-equilibrium system. We numerically illustrate these two inequalities on a model of biochemical enzyme reaction.

\textit{Stochastic thermodynamics.--} 
To clarify a link between stochastic thermodynamics and information geometry, we here start with the formalism of stochastic thermodynamics for the master equation~\cite{sekimoto2010stochastic, seifert2012stochastic}, that is also known as the Schnakenberg network theory~\cite{schnakenberg1976network, andrieux2007fluctuation}. 

We here consider a $(n+1)$-states system. We assume that transitions between states are induced by $n_{\rm bath}$-multiple thermal baths.
The master equation for the probability $p_x$ ($\geq 0$, $\sum_{x=0}^n p_x =1$) to find the state at $x=\{0, 1, \dots, n \}$ is given by
\begin{equation}  
\frac{d}{dt}p_x = \sum_{\nu =1}^{n_{\rm bath}} \sum_{x'=0}^n W^{(\nu)}_{x' \to x} p_{x'},
\label{eq:ME}
\end{equation} 
where $W^{(\nu)}_{x' \to x}$ is the transition rate from $x'$ to $x$ induced by $\nu$-th thermal bath. We assume a non-zero value of the transition rate $W^{(\nu)}_{x' \to x} >0$ for any $x \neq x'$. We also assume the condition
\begin{align}
\sum_{x=0}^{n} W^{(\nu)}_{x' \to x} &= 0,
\label{conservation}
\end{align} 
or equivalently $W^{(\nu)}_{x' \to x'} = - \sum_{x \neq x'} W^{(\nu)}_{x' \to x} <0$, which leads to the conservation of probability $  d(\sum_{x=0}^n p_x)/dt =0$.
This equation (\ref{conservation}) indicates that the master equation is then given by the thermodynamic flux from the state $x'$ to $x$~\cite{schnakenberg1976network},
\begin{align}
J_{x' \to x}^{(\nu)}:=W_{x' \to x}^{(\nu)} p_{x'} - W_{x \to x'}^{(\nu)} p_{x},
\end{align}
\begin{align}
\frac{d}{dt}p_x &= \sum_{\nu =1}^{n_{\rm bath}}   \sum_{x'=0}^n J^{(\nu)}_{x' \to x}.
\label{conti}
\end{align} 
If dynamics are reversible (i.e., $J^{(\nu)}_{x' \to x}=0$ for any $x$, $x'$ and $\nu$), the system is said to be in thermodynamic equilibrium. If we consider the conjugated thermodynamic force  
\begin{align}
F_{x' \to x}^{(\nu)} := \ln [W_{x' \to x}^{(\nu)} p_{x'} ]- \ln [W_{x \to x'}^{(\nu)} p_{x}],
\label{force}
\end{align} 
thermodynamic equilibrium is equivalently given by $F_{x' \to x}^{(\nu)}=0$ for any $x$, $x'$ and $\nu$. 

In stochastic thermodynamics~\cite{seifert2012stochastic}, we treat the entropy change of thermal bath and the system in a stochastic way. In the transition from $x'$ to $x$, the stochastic entropy change of $\nu$-th thermal bath is defined as
\begin{align}
\Delta {\sigma}^{{\rm bath (\nu)}}_{ x' \to x} &:= \ln \frac{W_{x' \to x}^{(\nu)}}{W_{x \to x'}^{(\nu)}},
 \label{bath}
\end{align}
and the stochastic entropy change of the system is defined as the stochastic Shannon entropy change
\begin{align}
\Delta {\sigma}^{\rm sys}_{ x' \to x} &:= \ln p_{x'}  - \ln p_x,
\end{align}  
respectively. 
The thermodynamic force is then given by the sum of entropy changes in the transition from $x'$ to $x$ induced by $\nu$-th thermal bath $F^{(\nu)}_{x' \to x} =\Delta {\sigma}^{{\rm bath}(\nu)}_{ x' \to x} +  \Delta{\sigma}^{\rm sys}_{x' \to x}$. This fact implies that the system is in equilibrium if the sum of entropy changes is zero for any transitions.

The total entropy production rate $\dot{\Sigma}^{\rm tot}$ is given by the sum of the products of thermodynamic forces and fluxes over possible transitions. To simplify notations, we introduce the set of directed edges $E= \{(x' \to x, \nu) | 0\leq x' < x \leq n , 1\leq \nu \leq n_{\rm bath}\}$ which denotes the set of all possible transitions between two states. The total entropy production rate is then given by
\begin{align}
\dot{\Sigma}^{\rm tot}&:=\sum_{(x' \to x, \nu) \in E} J^{(\nu)}_{x' \to x} F^{(\nu)}_{x' \to x}  =\langle F \rangle ,
\end{align} 
where a parenthesis $\langle \cdots \rangle$ is defined as $\langle A \rangle := \sum_{(x' \to x, \nu) \in E} J^{(\nu)}_{x' \to x} A^{(\nu)}_{x' \to x}$ for any function of edge $A^{(\nu)}_{x' \to x}$. Because signs of the thermodynamic force $F_{x' \to x}^{(\nu)}$ and the flux $J_{x' \to x}^{(\nu)}$ are same, the total entropy production rate is non-negative
\begin{align}
 \langle F \rangle =\langle \Delta {\sigma}^{{\rm bath}} \rangle + \langle \Delta {\sigma}^{{\rm sys}} \rangle \geq 0,
\end{align} 
that is well known as the second law of thermodynamics.

\textit{Information geometry.--} 
Next, we introduce information theory well known as information geometry~\cite{amari2007methods}. In this letter, we only consider the discrete distribution group ${\bf p} = (p_0, p_1, \dots, p_n)$, $p_x \geq 0$, and $\sum_{x=0}^n p_x =1$. This discrete distribution group gives the $n$-dimensional manifold $S_n$, because the discrete distribution is given by $n+1$ parameters $(p_0, p_1, \dots, p_n)$ under the constraint $\sum_{x=0}^n p_x =1$. To introduce a geometry on the manifold $S_n$, we conventionally consider the Kullback-Leibler divergence~\cite{cover2012elements} between two distributions ${\bf p}$ and ${\bf p}' =(p'_0, p'_1, \dots, p'_n)$ defined as
\begin{equation}
D_{\rm KL}({\bf p}|| {\bf p}' ) : = \sum_{x=0}^n p_x  \ln \frac{p_x}{p_x'}.
\end{equation} 
The square of the line element $ds$ is defined as the second-order Taylor series of the Kullback-Leibler divergence
\begin{equation}
ds^2 := \sum_{x=0}^n \frac{(dp_x)^2}{p_x} = 2 D_{\rm KL}({\bf p}|| {\bf p} + d{\bf p}), 
\label{metric}
\end{equation} 
where $d{\bf p} =  (dp_0, dp_1, \dots, dp_n)$ is the infinitesimal displacement that satisfies $\sum_{x=0}^n dp_x =0$. This square of the line element is directly related to the Fisher information metric~\cite{rao1992information} (see also Supplementary Information (SI)).
\begin{figure}[tb]
\centering
\includegraphics[width=5cm]{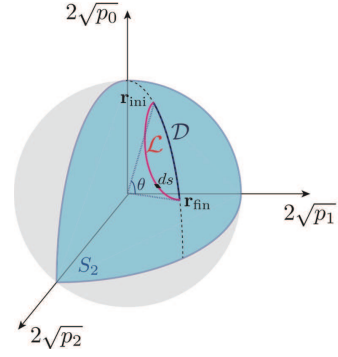}
\caption{(color online). Schematic of information geometry on the manifold $S_2$. The manifold $S_2$ leads to the sphere surface of radius $2$ (see also SI). The statistical length $\mathcal{L}$ is bounded by the shortest length $\mathcal{D} =2 \theta = 2 \cos^{-1} ({\bf r}_{\rm ini} \cdot {\bf r}_{\rm fin})$.}
\label{fig1}
\end{figure}

The manifold $S_n$ leads to the geometry of the $n$-sphere surface of radius $2$ (see also Fig. \ref{fig1}), because the square of the line element is also given by $ds^2 = \sum_{x=0}^n (2 dr_x)^2$ under the constraint ${\bf r} \cdot {\bf r}  =\sum_{x} (\sqrt{p_x})^2 = 1$ where ${\bf r}$ is the unit vector defined as ${\bf r}=(r_0, r_1, \dots, r_n) :=( \sqrt{p_0},\sqrt{p_1}, \dots, \sqrt{p_n})$ and $\cdot$ denotes the inner product. The statistical length $\mathcal L$~\cite{wootters1981statistical, braunstein1994statistical}
\begin{align}
{\mathcal L} := \int ds = \int \frac{ds}{dt} dt,
\end{align} 
from the initial state ${\bf r}_{\rm ini}$ to the final state ${\bf r}_{\rm fin}$ is then bounded by 
\begin{align}
{\mathcal L} & \geq  2 \cos^{-1} ({\bf r}_{\rm ini} \cdot {\bf r}_{\rm fin}) := \mathcal{D}({\bf r}_{\rm ini};{\bf r}_{\rm fin}),
\label{distanceineq}
\end{align}
because $\mathcal{D}({\bf r}_{\rm ini};{\bf r}_{\rm fin})=2 \theta$ is the shortest length between ${\bf r}_{\rm ini}$ and ${\bf r}_{\rm fin}$ on the $n$-sphere surface of radius $2$, where $\theta$ is the angle between ${\bf r}_{\rm ini}$ and ${\bf r}_{\rm fin}$ given by the inner product ${\bf r}_{\rm ini} \cdot {\bf r}_{\rm fin}=\cos \theta$.

\textit{Stochastic thermodynamics of information geometry.--} 
We here discuss a relationship between the line element and conventional observables of stochastic thermodynamics, which gives a stochastic thermodynamic interpretation of information geometric quantities.

By using the master equation (\ref{eq:ME}) and definitions of the line element and thermodynamic quantities Eqs.  (\ref{force}), (\ref{bath}) and (\ref{metric}), we obtain stochastic thermodynamic expressions of $ds^2/dt^2$ (see also SI), 
\begin{align}
\frac{ds^2}{dt^2} &= \sum_{x=0}^n p_x \frac{d}{dt} \left( -  \frac{1}{p_x} \frac {dp_x }{dt}  \right) \\ 
&= -\sum_{x=0}^n p_x \frac{d}{dt} \left( \sum_{\nu=1}^{n_{\rm bath}} \sum_{x'=0}^n W^{(\nu)}_{x \to x'}  e^{-F^{(\nu)}_{x \to x'} }  \right) \label{thermodynamics}\\
&= \left< \frac{d\Delta{\sigma}^{\rm bath}}{dt}\right>- \left< \frac{dF}{dt}\right>. 
\label{thermods}
\end{align}
Equation (\ref{thermodynamics}) implies that geometric dynamics are driven by the thermodynamic factor $\exp[ -F^{(\nu)}_{x \to x'} ]$, that is well discussed in the context of stochastic thermodynamics (especially in the context of the fluctuation theorem~\cite{jarzynski1997nonequilibrium, crooks1999entropy, evans2002fluctuation, seifert2005entropy, esposito2010three, esposito2010three2}). The time evolution of the line element  $ds^2/dt^2$ is directly related to the expected value of the time derivative of the rate-weighted thermodynamic factor $W^{(\nu)}_{x \to x'}  e^{-F^{(\nu)}_{x \to x'}}$.

Another expression Eq.~(\ref{thermods}) gives a stochastic thermodynamic interpretation of information geometry, especially in case of a near-equilibrium system. The condition of an equilibrium system is given by $F_{x' \to x}^{(\nu)} = 0$ for any $x'$, $x$ and $\nu$. Then, the square of the line element is given by the entropy change in thermal baths $ds^2 \simeq  \left< d\Delta{\sigma}^{\rm bath} \right> dt$ in a near-equilibrium system. 

For example, in a near-equilibrium system, the probability distribution is assumed to be the canonical distribution $p_x = \exp (\beta(\phi- H_x))$, where $\phi := - \beta^{-1} \ln[ \sum_{x=0}^n \exp (-\beta H_x))] $ is the Helmholtz free energy, $\beta$ is the inverse temperature and $H_x$ is the Hamiltonian of the system in the state $x$. To consider a near-equilibrium transition, we assume that $\beta$ and $H_x$ can depend on time. From $ds^2 =  [\left<d\Delta{\sigma}^{\rm bath}\right>- \left< dF\right>]dt = - \left< d\Delta{\sigma}^{\rm sys} \right> dt$, we obtain $ds^2 =-  \left< d\Delta{\sigma}^{\rm sys} \right> dt =  - \langle d ( \beta \Delta {H} )\rangle dt$ in a near equilibrium system, where $\Delta{H}_{x' \to x}:=H_{x} - H_{x'}$ is the Hamiltonian change from the state $x'$ to $x$. Because $-\beta \Delta{H}$ can be considered as the entropy change of thermal bath $\Delta{\sigma}^{\rm bath}$, an expression $ds^2 =  - \langle d (\beta \Delta{H} )\rangle dt$ for the canonical distribution is consistent with a near equilibrium expression $ds^2 \simeq  \left< d\Delta {\sigma}^{\rm bath} \right> dt$.

We also discuss the second order expansion of $ds^2/dt^2$ for the thermodynamic force in SI, based on the linear irreversible thermodynamics~\cite{schnakenberg1976network}. Our discussion implies that the square of the line element (or the Fisher information metric) for the thermodynamic forces is related to the Onsager coefficients. Due to the Cram\'{e}r-Rao bound~\cite{rao1992information, cover2012elements}, the Onsager coefficients are directly connected to a lower bound of the variance of unbiased estimator for parameters driven by the thermodynamic force.

Due to the non-negativity of the square of line element $ds^2/dt^2 \geq 0$, we have a thermodynamic inequality
\begin{align}
\left< \frac{d\Delta {\sigma}^{\rm bath}}{ dt} \right>  \geq \left< \frac{dF}{dt} \right>.
\label{ineqsub}
\end{align}
The equality holds if the system is in a stationary state, i.e., $dp_x/dt=0$ for any $x$. This result (\ref{ineqsub}) implies that the change of the thermodynamic force rate is transferred to the environmental entropy change rate. The difference $\langle d\Delta{\sigma}^{\rm bath}/dt \rangle - \langle dF/dt \rangle  \geq 0$ can be interpreted as loss in the entropy change rate transfer due to the non-stationarity. If the environmental entropy change does not change in time (i.e., $d\Delta {\sigma}^{\rm bath(\nu)}_{x' \to x}/dt =0$ for any $x'$ and $x$), the thermodynamic force change tends to decrease (i.e., $\langle dF/dt \rangle \leq 0$) in a transition. We stress that a mathematical property of the thermodynamic force in this result is different from the second law of thermodynamics $\langle F \rangle \geq 0$.

From Eq.~(\ref{thermods}), the statistical length $\mathcal{L} = \int^{\tau}_0 dt (ds/dt)$ from time $t=0$ to $t=\tau$ is given by 
\begin{align}
\mathcal{L} =\int_{t=0}^{t=\tau} dt \sqrt{\left< \frac{d\Delta{\sigma}^{\rm bath}}{dt}\right>- \left< \frac{dF}{dt}\right>}.
\label{length}
\end{align}
We then obtain the following thermodynamic inequality from Eqs.~(\ref{distanceineq}) and (\ref{length}), 
\begin{align} 
\int_{t=0}^{t=\tau} dt \sqrt{\left< \frac{d\Delta{\sigma}^{\rm bath}}{dt}\right>- \left< \frac{dF}{dt}\right>} \geq   \mathcal{D}({\bf r} (0);{\bf r} (\tau)).
\end{align}
The equality holds if the path of transient dynamics is a geodesic line on the manifold $S_n$. This inequality gives a geometric constraint of the entropy change rate transfer in a transition between two probability distributions ${\bf p} (0)$ and ${\bf p} (\tau)$. 

\begin{figure}[tb]
\centering
\includegraphics[width=8cm]{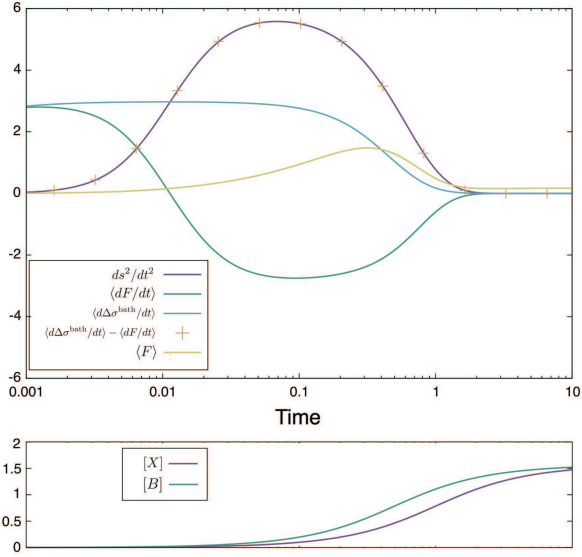}
\caption{(color online). Numerical calculation of thermodynamic quantities in the three states model of enzyme reaction. We numerically shows the non-negativity of $ds^2/dt^2 \geq 0$ and $ds^2/dt^2=- \langle dF/dt \rangle + \langle d\Delta {\sigma}^{\rm bath}/dt \rangle$ in the graph. We also show the total entropy change rate $\langle F \rangle \geq 0$. We note that $d\langle F \rangle/dt$ is not equal to $\langle dF/dt \rangle$.}
\label{fig2}
\end{figure}

\textit{Thermodynamic uncertainty.--} 
We finally reach to a thermodynamic uncertainty relationship between speed and thermodynamic cost. We here consider the action $\mathcal{C}:= (1/2) \int_{t=0}^{t=\tau} dt  (ds^2/dt^2)$ from time $t=0$ to $t=\tau$. From Eq.~(\ref{thermods}), the action $\mathcal{C}$ is given by
\begin{align}
\mathcal{C} &= \frac{1}{2} \int_{t=0}^{t=\tau} dt \left[\left< \frac{d\Delta{\sigma}^{\rm bath}}{dt}\right>- \left< \frac{dF}{dt}\right> \right].
\end{align}
Especillay in case of a near-equilibrium system, the action $\mathcal{C}$ is given by $\mathcal{C} \simeq \int \left< d\Delta{\sigma}^{\rm bath} \right>/2$. If we assume the canonical distribution, we have $\mathcal{C}  =  -\int \langle d (\beta \Delta{H} )\rangle/2$. Even for a system far from equilibrium, we can consider the action as a total amount of loss in the entropy change rate transfer. Therefore, the action can be interpreted as thermodynamic cost. 

Due to the Cauchy-Schwarz inequality $\int_0^{\tau} dt  \int_0^{\tau} (ds/dt)^2 dt \geq  (\int_0^{\tau} (ds/dt) dt)^2$~\cite{crooks2007measuring}, we obtain a thermodynamic uncertainty relationship between speed $\tau$ and thermodynamic cost $\mathcal{C}$
\begin{align}
 \tau \geq \frac{\mathcal{L}^2 }{2 \mathcal{C}}. 
\end{align}
The equality holds if speed of dynamics $ds^2/dt^2$ does not depend on time. By using the inequality (\ref{distanceineq}), we also have a weaker bound
\begin{align}
\tau \geq \frac{[\mathcal{D}({\bf r} (0);{\bf r} (\tau))]^2}{2 {\mathcal C}}.
\label{uncertainty}
\end{align}
In a transition from ${\bf r} (0)$ to ${\bf r} (\tau) (\neq {\bf r} (0))$, thermodynamic cost ${\mathcal C}$ should be large if the transition time $\tau$ is small. In case of a near-equilibrium system, we have $2\mathcal{C}  = \int \left< d\Delta{\sigma}^{\rm bath} \right>$ (or $2\mathcal{C}  =  -\int \langle d (\beta \Delta {H} )\rangle$), and then the inequality is similar to the quantum speed limit that is discussed in quantum mechanics~\cite{pires2016generalized}. We stress that this result is based on stochastic thermodynamics, not on quantum mechanics.

The inequality (\ref{uncertainty}) gives the ratio between time-averaged thermodynamic cost $2\mathcal{C}/ \tau$ and square of the velocity on manifold $([\mathcal{D}({\bf r} (0);{\bf r} (\tau))]/ \tau)^2$. Then, this ratio 
\begin{align}
\eta := \frac{[\mathcal{D}({\bf r} (0);{\bf r} (\tau))]^2}{2\tau {\mathcal C}}.
\label{efficiency}
\end{align}
quantifies an efficiency for power to speed conversion. Due to the inequality  (\ref{uncertainty}) and its non-negativity, the efficiency $\eta$ satisfies $0\leq \eta \leq 1$, where $\eta =1$ ($\eta =0$) implies high (low) efficiency.

\begin{figure}[tb]
\centering
\includegraphics[width=8cm]{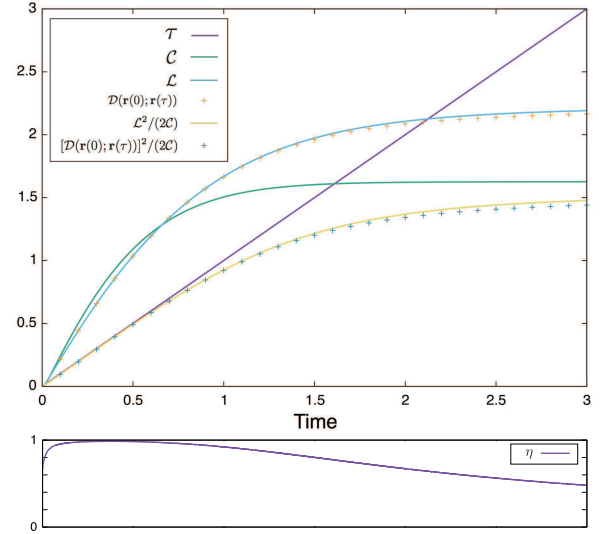}
\caption{(color online). Numerical calculation of the thermodynamic uncertainty relationship in the three states model of enzyme reaction. We numerically shows the geometric inequality $\mathcal{L} \geq \mathcal{D} ({\bf r} (0); {\bf r} (\tau))$, the thermodynamic uncertainty relationship $\tau \geq {\mathcal L}^2/ (2 \mathcal{C}) \geq  [{\mathcal D} ({\bf r} (0); {\bf r} (\tau))]^2/ (2 \mathcal{C})$, and the efficiency $\eta$ in the graph.}
\label{fig3}
\end{figure}

\textit{Three states model of enzyme reaction.--}  
We numerically illustrate thermodynamic inequalities of information geometry by using a thermodynamic model of biochemical reaction.
We here consider a three states model (see also SI) that represents a chemical reaction $A+B \rightleftharpoons AB$ with enzyme $X$,
\begin{align}
 A+X \rightleftharpoons AX, \\
 A+B \rightleftharpoons AB, \\
 AX +B \rightleftharpoons AB +X.
\end{align}
We here consider the probability distribution of states $x=A, AX, AB$. We assume that the system is attached to a single heat bath ($n_{\rm bath} =1$) with inverse temperature $\beta$. The master equation is given by Eq. (\ref{eq:ME}), where the transition rates are supposed to be
\begin{align}
W^{(1)}_{A \to AX} = k_{AX+} [X],&   &W^{(1)}_{AX \to A} = k_{AX+} e^{-\beta{\Delta \mu_{AX} }}, \nonumber\\
W^{(1)}_{A \to AB} = k_{AB+} [B], & &W^{(1)}_{AB \to A} = k_{AB+}e^{-\beta{\Delta \mu_{AB} }},  \nonumber\\
W^{(1)}_{AX \to AB} = k_{+} [B],& & W^{(1)}_{AB \to AX} = k_{+}e^{-\beta{\Delta \mu }} [X],
\end{align}
$[X]$ ($[B]$) is the concentration of $X$ ($B$), $k_{AX+}$, $k_{AB+}$, and $k_{+}$ are reaction rate constants, and ${\Delta \mu_{AX} }$, ${\Delta \mu_{AB} }$, and $\Delta \mu$ are the chemical potential differences. In this model, the entropy change of bath $\Delta {\sigma}^{{\rm bath (\nu)}}_{ x' \to x}$ is given by this chemical potential difference (see also SI)~\cite{schmiedl2007stochastic}.

In a numerical simulation, we set $k_{AX+} =k_{AB+} =k_+ =1$, $\beta \Delta \mu_{AX}= 1$,  $\beta \Delta \mu_{AB} = 0.5$, and $\beta \Delta \mu= 2$. We assume that the time evolution of the concentrations is given by $[X] = \tan^{-1}(\omega_X t)$, $[B] = \tan^{-1}(\omega_B t)$ with $\omega_X=1$ and $\omega_B=2$, which means that the concentrations $[X]$ and $[B]$ perform as control parameters. At time $t=0$, we set the initial probability distribution as $(p_A, p_{AX},p_{AB})=(0.9998,0.0001,0.0001)$.

In Fig.~\ref{fig2}, we numerically show the inequality $\langle d\Delta{\sigma}^{\rm bath}/dt \rangle \geq \langle dF/dt \rangle$. We check that this inequality does not coincide with the second law of thermodynamics $\langle F \rangle \geq 0$. We also check the thermodynamic uncertainty relationship $\tau \geq \mathcal{L}^2 / (2\mathcal{C})$ in Fig.~\ref{fig3}. Because the path from the initial distribution $(p_A, p_{AX},p_{AB})=(0.9998,0.0001,0.0001)$ to the final distribution is close to the geodesic line, the thermodynamic uncertainty relationship gives a tight bound of the transition time $\tau$.

\textit{Conclusion.--} 
In this letter, we reveal a link between stochastic thermodynamic quantities ($J$, $F$, $\Delta{\sigma}^{\rm sys}$, $\Delta{\sigma}^{\rm bath}$) and information geometric quantities ($ds^2$, $\mathcal{L}$, $\mathcal{D}$, $\mathcal{C}$). Because the theory of information geometry is applicable to various fields of science such as neuroscience, signal processing, machine learning and quantum mechanics, this link would help us to understand a thermodynamic aspect of such a topic. The trade-off relationship between speed and thermodynamic cost Eq. (\ref{uncertainty}) would be helpful to understand biochemical reactions and gives a new insight into recent studies of the relationship between information and thermodynamics in biochemical processes~\cite{ito2013information, barato2014efficiency, sartori2014thermodynamic, ito2015maxwell, bo2015thermodynamic, ouldridge2017thermodynamics, mcgrath2017biochemical}.

\section{acknowledgement}
I am grateful to Shumpei Yamamoto for discussions of stochastic thermodynamics for the master equation, to Naoto Shiraishi, Keiji Saito, Hal Tasaki, and Shin-Ichi Sasa for discussions of thermodynamic uncertainty relationships, to Schuyler B. Nicholson for discussion of information geometry and thermodynamics, and to Pieter rein ten Wolde for discussions of thermodynamics in a chemical reaction.  We also thank Tamiki Komatsuzaki to acknowledge my major contribution of this work and allow me to submit this manuscript alone. I mentioned that, after my submission of the first version of this manuscript on arXiv~\cite{SosukeIto}, I heard that Schuyler B. Nicholson independently discovered a similar result such as Eq. (\ref{thermods})~\cite{Schuyler}. I thank Sebastian Goldt, Matteo Polettini, Taro Toyoizumi, and Hiroyasu Tajima for valuable comments on the manuscript. This research is supported by JSPS KAKENHI Grant No. JP16K17780.

\widetext

\section{Supplementary information}
\subsection{I. Intuitive proof of the fact that the manifold $S_2$ gives the sphere surface of radius $2$}
We here intuitively show the fact that the manifold $S_2$ gives the sphere surface of radius $2$. The set of probability ${\bf p} = (p_0, p_1, p_2)$ satisfies the normalization $\sum_{x=0}^2 p_x =1$. The square of the line element $ds$ is given by
\begin{align}
ds^2 &= \sum_{x=0}^2 \frac{(dp_x)^2}{p_x}.
\label{suppds}
\end{align}
We here introduce the polar coordinate system $(\phi, \psi)$ where $p_0 = (\cos \psi)^2$, $p_1 = (\sin \psi)^2 (\cos \phi)^2$, $p_2 = (\sin \psi)^2 (\sin \phi)^2$. We can check that the normalization $\sum_{x=0}^2 p_x =1$ holds. 
By using the polar coordinate system, $(dp_0, dp_1, dp_2)$ is given by $dp_0 = -2(\cos \psi)(\sin \psi)d\psi$,  $dp_1 = 2(\cos \psi)(\sin \psi)  (\cos \phi)^2d\psi - 2(\cos \phi)(\sin \phi)  (\sin \psi)^2d\phi$, and $dp_2 = 2(\cos \psi)(\sin \psi)  (\sin \phi)^2d\psi + 2(\cos \phi)(\sin \phi)  (\sin \psi)^2d\phi$. From Eq. (\ref{suppds}), we then obtain
\begin{align}
ds^2 &= 4[ (\sin \psi)^2 + (\cos \psi)^2 (\cos \phi)^2 + (\cos \psi)^2 (\sin \phi)^2](d\psi)^2 +0\times (d\phi)(d\psi)+ 4 [(\sin \phi)^2  (\sin \psi)^2 + (\cos \phi)^2 (\sin \psi)^2] (d\phi)^2 \nonumber\\
&= 2^2 [(d\psi)^2 + (\sin \psi)^2(d\phi)^2 ].
\end{align}
Because the metric of the sphere surface of radius $R$ is given by $ds^2 = R^2 [(d\psi)^2 + (\sin \psi)^2(d\phi)^2 ]$, the manifold $S_2$ gives the sphere surface of radius $R=2$. 

\subsection{II. Detailed derivation of Eqs. (15) and (16) in the main text}
We here discuss the detailed derivation of Eqs. (15) and (16) in the main text, and the relationship between the square of the line element and the Fisher information metric. 

By using the definition of the thermodynamic force $F_{x' \to x}^{(\nu)} := \ln [W_{x' \to x}^{(\nu)} p_{x'} ]- \ln [W_{x \to x'}^{(\nu)} p_{x}]$,  the master equation is given by
\begin{align}
\frac{d}{dt}p_x &= \sum_{\nu=1}^{n_{\rm bath}}\sum_{x'=0}^n W^{(\nu)}_{x \to x'} p_{x} e^{-F^{(\nu)}_{x \to x'} }.
\label{supmasterthermo}
\end{align}
From Eqs. (\ref{suppds}), (\ref{supmasterthermo}) and $\sum_{x=0}^n d^2p_x /dt^2  = 0$, we obtain an expression Eq. (15) in the main text,
\begin{align}
\frac{ds^2}{dt^2} &= \sum_{x=0}^n \frac{1}{p_x} \left( \frac{dp_x}{dt} \right)^2 \nonumber\\ 
&= \sum_{x=0}^n p_x \frac{d}{dt} \left( -\frac{1}{p_x} \right) \left( \frac{dp_x}{dt} \right)   \nonumber\\ 
&= \sum_{x=0}^n p_x \frac{d}{dt} \left( -\frac{1}{p_x} \right) \left( \frac{dp_x}{dt} \right) - \sum_{x=0}^n \left( \frac{d^2p_x}{dt^2} \right)   \nonumber\\ 
&= \sum_{x=0}^n p_x \frac{d}{dt} \left( -  \frac{1}{p_x} \frac {dp_x }{dt}  \right) \nonumber\\ 
&= -\sum_{x=0}^n p_x \frac{d}{dt} \left( \sum_{\nu=1}^{n_{\rm bath}} \sum_{x'=0}^n W^{(\nu)}_{x \to x'}  e^{-F^{(\nu)}_{x \to x'} }  \right).
\label{supthermodynamics}
\end{align}
Let ${\mathbb E}[ A] := \sum_{x=0}^n p_x A(x)$ be the expected value of any function $A(x)$,  and $\overline{A} (x) :=\sum_{\nu=1}^{n_{\rm bath}} \sum_{x'=0}^n W^{(\nu)}_{x \to x'} A^{(\nu)}_{x \to x'}$ be the rate-weighted expected value of any function of edge $A^{(\nu)}_{x \to x'}$ with a fixed initial state $x$, respectively. By using these notation, the result (\ref{supthermodynamics}) can be rewritten as
\begin{align}
\frac{ds^2}{dt^2} = -{\mathbb E} \left[ \frac{d}{dt} \overline{e^{-F } }  \right].
\end{align}
We here mention that a parenthesis in the main text is given by $\langle A \rangle = {\mathbb E} \left[ \overline{A} \right]$ if $A^{(\nu)}_{x \to x'}$ is an anti-symmetric function $A^{(\nu)}_{x \to x'} = - A^{(\nu)}_{x' \to x}$. Because the thermodynamic force is an anti-symmetric function $F^{(\nu)}_{x \to x'} = - F^{(\nu)}_{x' \to x}$,  the total entropy production rate is given by $\dot{\Sigma}^{\rm tot} = {\mathbb E} \left[ \overline{F}\right]$. We also carefully mention that the expected value of $\overline{e^{-F}}$ gives ${\mathbb E} [\overline{e^{-F}}]=\sum_{\nu=1}^{n_{\rm bath}} \sum_{x=0}^{n} \sum_{x'=0}^{n} p_{x'} W^{(\nu)}_{x' \to x} =0$, compared to the integral fluctuation theorem $\langle e^{-F_{\rm traj}} \rangle_{\rm traj} =1$ with the entropy production of trajectories $F_{\rm traj}$ and the ensemble average of trajectories $\langle \cdots \rangle_{\rm traj}$~\cite{supseifert2005entropy, supesposito2010three2}. If the system is in a stationary state, i.e., $dp_x/ dt=0$ for any $x$, we have 
\begin{align}
{\mathbb E} \left[ \frac{d}{dt} \overline{e^{-F } }  \right] =\frac{d}{dt} \left( {\mathbb E} \left[ \overline{e^{-F } }  \right] \right) =0.
\end{align}

From Eq. (\ref{supthermodynamics}), we also obtain 
\begin{align}
\frac{ds^2}{dt^2} =&  -\sum_{x=0}^n p_x \frac{d}{dt} \left( \sum_{\nu=1}^{n_{\rm bath}} \sum_{x'=0}^n W^{(\nu)}_{x \to x'}  e^{-F^{(\nu)}_{x \to x'} }  \right) \nonumber \\
=& -\sum_{x=0}^n p_x \left(\sum_{\nu=1}^{n_{\rm bath}}  \sum_{x'=0}^n W^{(\nu)}_{x \to x'} \left(-\frac{d}{dt}F^{(\nu)}_{x \to x'} \right)e^{-F^{(\nu)}_{x \to x'} }  \right) -\sum_{x=0}^n p_x \left( \sum_{\nu=1}^{n_{\rm bath}}  \sum_{x'=0}^n \left(\frac{d}{dt}W^{(\nu)}_{x \to x'} \right) e^{-F^{(\nu)}_{x \to x'} }  \right).
\end{align}
The first term is calculated as follows
\begin{align}
&-\sum_{x=0}^n p_x \left(\sum_{\nu=1}^{n_{\rm bath}}  \sum_{x'=0}^n W^{(\nu)}_{x \to x'} \left(-\frac{d}{dt}F^{(\nu)}_{x \to x'} \right)e^{-F^{(\nu)}_{x \to x'} }  \right)  \nonumber \\
=&  -\sum_{\nu=1}^{n_{\rm bath}} \sum_{x=0}^n  \sum_{x'=0}^n  p_{x'} W^{(\nu)}_{x' \to x}\left(\frac{d}{dt}F^{(\nu)}_{x' \to x} \right)   \nonumber \\
=& -\sum_{\nu=1}^{n_{\rm bath}} \sum_{x, x'| x>x'} p_{x'} W^{(\nu)}_{x' \to x}\left(\frac{d}{dt}F^{(\nu)}_{x' \to x} \right)  - \sum_{\nu=1}^{n_{\rm bath}} \sum_{x, x'| x'>x}  p_{x'} W^{(\nu)}_{x' \to x}\left(\frac{d}{dt}F^{(\nu)}_{x' \to x} \right) \nonumber \\
=& -\sum_{(x' \to x, \nu) \in E} J^{(\nu)}_{x' \to x} \left(\frac{d}{dt}F^{(\nu)}_{x' \to x} \right)= - \left< \frac{dF}{dt}\right>,
\end{align}
where we used $F^{(\nu)}_{x' \to x} = -F^{(\nu)}_{x \to x'}$ and $F^{(\nu)}_{x' \to x'} = 0$.
The second term is also calculated as follows
\begin{align}
&-\sum_{x=0}^n p_x \left( \sum_{\nu=1}^{n_{\rm bath}}  \sum_{x'=0}^n \left(\frac{d}{dt}W^{(\nu)}_{x \to x'} \right) e^{-F^{(\nu)}_{x \to x'} }  \right) \nonumber \\
=&  -\sum_{\nu=1}^{n_{\rm bath}}  \sum_{x=0}^n  \sum_{x'=0}^n  p_{x'} W^{(\nu)}_{x' \to x} \frac{1}{W^{(\nu)}_{x \to x'}}\left(\frac{d}{dt}W^{(\nu)}_{x \to x'} \right)  \nonumber \\
=& - \sum_{\nu=1}^{n_{\rm bath}}  \sum_{x, x'| x' \neq x}  p_{x'} W^{(\nu)}_{x' \to x} \frac{1}{W^{(\nu)}_{x \to x'}}\left(\frac{d}{dt}W^{(\nu)}_{x \to x'} \right)  - \sum_{\nu=1}^{n_{\rm bath}}\sum_{x=0}^n p_{x} W^{(\nu)}_{x \to x} \frac{1}{W^{(\nu)}_{x \to x}}\left(\frac{d}{dt}W^{(\nu)}_{x \to x} \right)  \nonumber \\
=& -\sum_{\nu=1}^{n_{\rm bath}}\sum_{x, x'| x>x'} p_{x'} W^{(\nu)}_{x' \to x} \frac{1}{W^{(\nu)}_{x \to x'}}\left(\frac{d}{dt}W^{(\nu)}_{x \to x'} \right)   + \sum_{\nu=1}^{n_{\rm bath}}\sum_{x=0}^n p_{x} \left( \sum_{x' \neq x}  \frac{d}{dt}W^{(\nu)}_{x \to x'} \right)  \nonumber \\
=& -\sum_{\nu=1}^{n_{\rm bath}} \sum_{x, x'| x \neq x'} p_{x'} W^{(\nu)}_{x' \to x} \left(\frac{d}{dt} \ln (W^{(\nu)}_{x \to x'}) \right)  + \sum_{\nu=1}^{n_{\rm bath}}\sum_{x, x' |x' \neq x}  p_{x'} W^{(\nu)}_{x' \to x} \left( \frac{d}{dt} \ln (W^{(\nu)}_{x' \to x} )\right)  \nonumber \\
=& \sum_{\nu=1}^{n_{\rm bath}} \sum_{x, x' |x' \neq x }  p_{x'} W^{(\nu)}_{x' \to x} \left( \frac{d}{dt} \Delta {\sigma}^{{\rm bath}(\nu)}_{x' \to x}\right)  \nonumber \\
=& \sum_{\nu=1}^{n_{\rm bath}} \sum_{x, x'| x>x'} p_{x'} W^{(\nu)}_{x' \to x}\left(\frac{d}{dt} \Delta {\sigma}^{{\rm bath}(\nu)}_{x' \to x} \right)- \sum_{\nu=1}^{n_{\rm bath}} \sum_{x, x'| x>x'}  p_{x} W^{(\nu)}_{x \to x'}\left(\frac{d}{dt} \Delta {\sigma}^{{\rm bath}(\nu)}_{x \to x'} \right) \nonumber \\
=& \sum_{(x' \to x, \nu) \in E} J^{(\nu)}_{x' \to x} \left(\frac{d}{dt} \Delta {\sigma}^{{\rm bath}(\nu)}_{x' \to x} \right) = \left< \frac{d\Delta {\sigma}^{\rm bath}}{dt}\right>,
\end{align}
where we used $W^{(\nu)}_{x' \to x'} = - \sum_{x \neq x'} W^{(\nu)}_{x' \to x}$, $ \Delta {\sigma}^{{\rm bath}(\nu)}_{x' \to x} = - \Delta {\sigma}^{{\rm bath}(\nu)}_{x \to x'}$ and $ \Delta {\sigma}^{{\rm bath}(\nu)}_{x' \to x'} = 0$.

By using $F^{(\nu)}_{x' \to x} = \Delta {\sigma}^{{\rm bath}(\nu)}_{x' \to x'}  + \Delta {\sigma}^{\rm sys}_{x' \to x'}$, we obtain an expression
\begin{align}
\frac{ds^2}{dt^2} =\left< \frac{d\Delta {\sigma}^{\rm bath}}{dt}\right> - \left< \frac{dF}{dt}\right> =  -\left< \frac{d\Delta {\sigma}^{\rm sys}}{dt}\right>.
\label{suppAA}
\end{align}

Let $(\lambda_1, \dots, \lambda_{n'})$ be the set of parameters such as control parameters. 
We also obtain the definition of the Fisher information metric~\cite{supcover2012elements} 
\begin{align}
g_{ij} =  \mathbb{E} \left[ \left( \frac{\partial \ln p}{\partial  \lambda_i} \right) \left(\frac{\partial \ln p}{\partial  \lambda_j} \right) \right] = \sum_{x=0}^n p_x  \left[ \left( \frac{\partial \ln p_x}{\partial  \lambda_i} \right) \left(\frac{\partial \ln p_x}{\partial  \lambda_j} \right) \right]
\end{align} from the result (\ref{suppAA}), 
\begin{align}
\frac{ds^2}{dt^2} =& \left< \frac{d\Delta {\sigma}^{\rm bath}}{dt}\right> - \left< \frac{dF}{dt}\right>  \nonumber\\
=& - \sum_{(x' \to x, \nu) \in E} J^{(\nu)}_{x' \to x} \left[ \frac{1}{p_{x'}} \frac{dp_{x'}}{dt} -  \frac{1}{p_{x}} \frac{dp_{x}}{dt} \right] \nonumber \\
=& - \sum_{\nu=1}^{n_{\rm bath}} \sum_{\nu'=1}^{n_{\rm bath}} \sum_{x=0}^n \sum_{x'=0}^n \sum_{x''=0}^n p_{x'}W^{(\nu)}_{x' \to x} \left[ \frac{1}{p_{x'}}  W^{(\nu')}_{x'' \to x'} p_{x''} -   \frac{1}{p_{x}} W^{(\nu')}_{x'' \to x} p_{x''}  \right] \nonumber \\
=& \sum_{x=0}^n p_x \left[ \sum_{\nu=1}^{n_{\rm bath}}  \sum_{x'=0}^n \frac{p_{x'}W^{(\nu)}_{x' \to x}}{p_x} \sum_{\nu'=1}^{n_{\rm bath}} \sum_{x''=0}^n \frac{p_{x''} W^{(\nu')}_{x'' \to x} }{p_{x}}  \right] \nonumber \\
=&  \sum_{x=0}^n p_x \left(\frac{d\ln p_x}{dt}\right)^2 \nonumber \\
=&  \sum_{x=0}^n p_x \left(\sum_{i=1}^{n'} \frac{\partial \ln p_x}{\partial \lambda_i} \frac{d\lambda_i}{dt} \right)^2  \nonumber \\
=& \sum_{i=1}^{n'} \sum_{j=1}^{n'} \frac{d\lambda_i}{dt} g_{ij} \frac{d\lambda_j}{dt},
\end{align}
where we used $\sum_{x=0}^n W^{(\nu)}_{x' \to x} =0$ and the master equation $dp_x /dt = \sum_{\nu=1}^{n_{\rm bath}}  \sum_{x'=0}^n p_{x'} W^{(\nu)}_{x' \to x}$. This result is consistent with the following calculation about the Fisher information metric
\begin{align}
ds^2 = \sum_{x=0}^n p_x (d\ln p_x)^2 = \sum_{x=0}^n p_x \left[ \sum_{i=1}^{n'} \left(\frac{\partial \ln p_x}{\partial \lambda_i}  \right) d\lambda_i \right]^2 = \sum_{i=1}^{n'} \sum_{j=1}^{n'}g_{ij} d\lambda_i d\lambda_j.
\end{align}

\subsection{III. Linear irreversible thermodynamic interpretation of information geometry}

We here discuss a stochastic thermodynamic interpretation of information geometry in a near-equilibirum system, where the entropy production rate is given by the second order expansion for the thermodynamic flow (or the thermodynamic force). This second order expansion is well known as linear irreversible thermodynamics~\cite{supschnakenberg1976network}. 

If we assume $F^{(\nu)}_{x' \to x} = 0$, we have $J^{(\nu)}_{x' \to x} =  0$. Thus, we have a linear expansion of thermodynamic force $F^{(\nu)}_{x' \to x}$ in terms of the thermodynamic flow $J^{(\nu)}_{x' \to x}$ for a near-equilibrium condition (i.e., $F^{(\nu)}_{x' \to x} \simeq 0$ for any $x$ and $x'$)
\begin{align}
F^{(\nu)}_{x' \to x}  &= \ln \left(1+ \frac{J^{(\nu)}_{x' \to x} }{W^{(\nu)}_{x' \to x} p_x }\right) \nonumber \\
 &=  \alpha^{(\nu)}_{x' \to x} J^{(\nu)}_{x' \to x} + o (J^{(\nu)}_{x' \to x}), \\
\alpha^{(\nu)}_{x' \to x} &:= \left. \frac{1}{W^{(\nu)}_{x' \to x} p_x } \right|_{ F^{(\nu)}_{x' \to x}=0}.
\end{align} 
We call this coefficient $\alpha^{(\nu)}_{x' \to x}$ as the Onsager coefficient of the edge $( x' \to x, \nu )$. The symmetry of the coefficient $\alpha^{(\nu)}_{x' \to x} =\alpha^{(\nu)}_{x \to x'}$ holds due to the condition $F^{(\nu)}_{x' \to x}=0$. 

If we consider the Kirchhoff's current law in a stationary state, the linear combination of the coefficient $\alpha^{(\nu)}_{x \to x'}$ leads to the Onsager coefficient~\cite{supschnakenberg1976network}. Let $\{C_1, \dots, C_m \}$ be the cycle basis of the Markov network for the master equation. The thermodynamic force of the cycle $F(C_i)$ is defined as
\begin{align}
F(C_i)= \sum_{(x' \to x, \nu )\in E} S(\{x' \to x, \nu\}, C_i) F_{x' \to x}^{(\nu)}
\end{align} 
where 
\begin{align}
  S(\{x' \to x, \nu\}, C_i) = \begin{cases}
    1 & (\{x' \to x, \nu\} \in C_i) \\
    -1 & (\{x \to x', \nu\} \in C_i) \\
    0 & ({\rm otherwise})
  \end{cases}.
\end{align}
The thermodynamic flow of the cycle $J(C_i)$ is defined as 
\begin{align}
J_{x' \to x}^{(\nu)}= \sum_{i=1}^m S(\{x' \to x, \nu \}, C_i) J(C_i).
\end{align} 
 We then obtain the linear relationship $F(C_j) =\sum_{i=1}^m L_{ji} J(C_i)$ (or $J(C_j) =\sum_{i=1}^m L^{-1}_{ji} F(C_i)$) with the Onsager coefficient 
 \begin{align}
L_{ij} =\sum_{(x' \to x, \nu) \in E}  \alpha_{x' \to x}^{(\nu)}  S(\{x' \to x, \nu\}, C_i)  S(\{x' \to x, \nu\}, C_j),
\end{align}
for a near-equilibrium condition, the second law of thermodynamics 
 \begin{align}
0 &\leq \dot{\Sigma}^{\rm tot} \nonumber \\
&= \sum_{(x' \to x, \nu )\in E} J_{x' \to x}^{(\nu)} F_{x' \to x}^{(\nu)} \nonumber \\
&= \sum_{(x' \to x, \nu )\in E} \sum_{i=1}^m S(\{x' \to x, \nu \}, C_i) J(C_i) F_{x' \to x}^{(\nu)} \nonumber \\
&= \sum_{i=1}^m J(C_i) F (C_i), \nonumber \\
&= \sum_{j=1}^m \sum_{i=1}^m L_{ij} J(C_i) J(C_j), \nonumber \\
&= \sum_{j=1}^m \sum_{i=1}^m L^{-1}_{ij} F(C_i) F(C_j), \nonumber \\
\end{align} 
and the Onsager reciprocal relationship $L_{ij} =L_{ji}$. This result gives the second order expansion of the entropy production rate $\dot{\Sigma}^{\rm tot}$ for the thermodynamic flow $J$ (or the thermodynamic force $F$) in a stationary state. For $m=2$, the second law of thermodynamics $L_{11} F(C_1)^2 + L_{22} F(C_2)^2 + 2L_{12} F(C_1)F(C_2) \geq 0$ is then given by $L_{11} \geq 0$, $L_{22} \geq 0$, and $L_{11}L_{22} -L_{12}^2 \geq 0$. 

Here we newly consider the second order expansion of $ds^2$ for the thermodynamic flow $J$ (or the thermodynamic force $F$) in linear irreversible thermodynamics. In a near-equilibrium system, the square of line element $ds$ is calculated as follows
\begin{align}
ds^2 =& - \left< \frac{d\Delta {\sigma}^{\rm sys}}{dt}\right>dt^2  \nonumber\\
=& - \sum_{(x' \to x, \nu )\in E} J^{(\nu)}_{x' \to x} \left[ \frac{1}{p_{x'}} \frac{dp_{x'}}{dt} -  \frac{1}{p_{x}} \frac{dp_{x}}{dt} \right] dt^2   \nonumber \\
=& - \sum_{(x' \to x, \nu) \in E} J^{(\nu)}_{x' \to x} \left[ \frac{1}{p_{x'}}  \sum_{\nu' =1}^{n_{\rm bath}}   \sum_{x''=0}^n J^{(\nu')}_{x'' \to x'}-  \frac{1}{p_{x}}  \sum_{\nu' =1}^{n_{\rm bath}}   \sum_{x''=0}^n J^{(\nu')}_{x'' \to x}  \right] dt^2 \nonumber \\
=& \sum_{(x' \to x, \nu) \in E}  \sum_{\nu' =1}^{n_{\rm bath}} \sum_{x''=0}^n J^{(\nu)}_{x' \to x} \left[  \frac{1}{p_{x}} J^{(\nu')}_{x'' \to x} - \frac{1}{p_{x'}}  J^{(\nu')}_{x'' \to x'}  \right]dt^2  \nonumber \\
=& \sum_{(x' \to x, \nu) \in E}  \sum_{\nu' =1}^{n_{\rm bath}} \sum_{x''=0}^n J^{(\nu)}_{x' \to x} \left[  \frac{1}{p_{x}} J^{(\nu')}_{x'' \to x} - \frac{1}{p_{x'}}  J^{(\nu')}_{x'' \to x'}  \right] dt^2 \nonumber \\
=& \sum_{\nu =1}^{n_{\rm bath}}  \sum_{\nu' =1}^{n_{\rm bath}} \sum_{x=0}^n \sum_{x'=0}^n \sum_{x''=0}^n \left[  \frac{J^{(\nu)}_{x' \to x }J^{(\nu')}_{x'' \to x}}{p_{x}}  \right]dt^2  \nonumber \\
=& \sum_{\nu =1}^{n_{\rm bath}}  \sum_{\nu' =1}^{n_{\rm bath}} \sum_{x=0}^n \sum_{x'=0}^n \sum_{x''=0}^n \left[  \frac{F^{(\nu)}_{x' \to x }F^{(\nu')}_{x'' \to x}}{\alpha^{(\nu)}_{x' \to x}  p_{x} \alpha^{(\nu')}_{x'' \to x} }  \right]dt^2  .
\end{align}
We here consider the situation that the time evolution of control parameters $\lambda_{(x' , x , \nu_x)}$ is driven by the thermodynamic force $F^{(\nu_x)}_{x' \to x} =d \lambda_{(x' , x , \nu_x)}/dt$. The square of line element can be written by the following Fisher information metric 
\begin{align}
ds^2 =& \sum_{\nu_x =1}^{n_{\rm bath}}  \sum_{x=0}^n \sum_{x'=0}^n  \sum_{\nu_y =1}^{n_{\rm bath}} \sum_{y=0}^n  \sum_{y'=0}^n g_{(x',x, \nu_x) (y', y, \nu_y)}d \lambda_{(x' , x , \nu_x)} d \lambda_{(y' , y , \nu_y) } \\
g_{(x',x, \nu_x) (y', y, \nu_y)} =& \frac{\delta_{xy}}{\alpha^{(\nu_x)}_{x' \to x}  p_{x} \alpha^{(\nu_y)}_{y' \to y} }.
\end{align}
This result implies that the Fisher information metric for control parameters $\lambda_{(x' , x , \nu_x)}$ driven by the thermodynamic force $F^{(\nu_x)}_{x' \to x} =d \lambda_{(x' , x , \nu_x)}/dt$ is related to the Onsager coefficients of the edge $\alpha^{(\nu_x)}_{x' \to x}$ for a near-equilibrium condition. Because the Cram\'{e}r-Rao bound~\cite{suprao1992information, supcover2012elements} implies that the variance of unbiased estimator is bounded by the inverse of this Fisher information metric, the Onsager coefficients of the edge $\alpha^{(\nu_x)}_{x' \to x}$ gives a lower bound of the variance of unbiased estimator for control parameters  driven by the thermodynamic forces in a near-equilibrium system.

\subsection{IV. Detail of the three states model of enzyme reaction}
Stochastic thermodynamics for the master equation is applicable to a model of chemical reaction~\cite{schmiedl2007stochastic}. 
We here discuss the thermodynamic detail of the three states model of enzyme reaction discussed in the main text. 

The master equation for Eq. (27) in the main text is given by
\begin{align}
\frac{dp_A}{dt} & =-(k_{AX+}[X] +k_{AB+}[B])  p_A + k_{AB-}p_{AB} +k_{AX-} p_{AX},\nonumber\\
\frac{dp_{AB}}{dt} &=  k_{AB+}[B]p_A-(k_{AB-} +k_{-} [X])p_{AB} +k_{+} [B] p_{AX} , \nonumber\\
\frac{dp_{AX}}{dt} &=  k_{AX+} [X] p_A+ k_{-}  [X] p_{AB} -(k_{AX-}+k_{+} [B]) p_{AX},
\label{supmasterchem}
\end{align}
where $k_{-}$, $k_{AB-}$ and $k_{AX-}$ are given by the chemical potential differences
\begin{align}
\ln \frac{k_{AX+}}{k_{AX-}} &= \beta \Delta \mu_{AX},  \nonumber\\
\ln \frac{k_{AB+}}{k_{AB-}} &= \beta{\Delta \mu_{AB} },  \nonumber\\
\ln \frac{k_{+}}{k_{-}} &= \beta{\Delta \mu}.
\end{align}

We here assume that the sum of the concentrations $[A]+ [AB]+[AX]=n_{A}$ is constant. The probabilities distributions $p_A$, $p_{AB}$, and $p_{AX}$ correspond to the fractions of $p_A=[A]/n_A$, $p_{AB}=[AB]/n_A$ and $p_{AX}=[AX]/n_A$, respectively. From the master equation (\ref{supmasterchem}), we obtain the rate equations of enzyme reaction
\begin{align}
\frac{d[A]}{dt} & =-(k_{AX+} [X]+k_{AB+}[B])  [A] + k_{AB-}[AB] +k_{AX-} [AX],\nonumber\\
\frac{d[AB]}{dt} &=  k_{AB+}[B][A]-(k_{AB-} +k_{-} [X])[AB] +k_{+} [B] [AX] , \nonumber\\
\frac{d[AX]}{dt} &=  k_{AX+} [X][A]+ k_{-} [X][AB] -(k_{AX-}+k_{+} [B]) [AX].
\end{align}
which corresponds to the following enzyme reaction
\begin{align}
 A+X \rightleftharpoons AX, \nonumber\\
 A+B \rightleftharpoons AB, \nonumber\\
 AX +B \rightleftharpoons AB +X,
\end{align}
where $A$ is substrate, $X$ is enzyme, $AX$ is enzyme-substrate complex, and $AB$ is product. 

In this model, the stochastic entropy changes of thermal bath are also calculated as
\begin{align}
\Delta {\sigma}^{{\rm bath}(1)}_{A \to AB}& = \beta \Delta \mu_{AB}+\ln [B],\nonumber\\
\Delta {\sigma}^{{\rm bath}(1)}_{AB \to AX}& = -\beta \Delta \mu -\ln [B] + \ln [X],\nonumber\\
\Delta {\sigma}^{{\rm bath}(1)}_{AX \to A}& = -\beta \Delta \mu_{AX} -\ln [X],
\end{align}
which are the conventional definitions of the stochastic entropy changes of a thermal bath.  In this model, the cycle basis is given by one cycle $\{C_1 =(A \to AB \to AX \to A)\}$. If the chemical potential change in a cycle $C_1$ has non-zero value, i.e., $\Delta \mu_{\rm cyc} :=\Delta \mu_{AB} - \Delta \mu - \Delta \mu_{AX} \neq 0$, the system in a stationary state is driven by the thermodynamic force of the cycle $F(C_1) = F^{(1)}_{A \to AB} + F^{(1)}_{AB \to AX} + F^{(1)}_{AX \to A} = \beta \Delta \mu_{\rm cyc}$. 
 In a numerical calculation, we set $ \beta \Delta \mu_{\rm cyc} =-2.5\neq 0$. Then we consider non-equilibrium and non-stationary dynamics in a numerical calculation.

\end{document}